\documentclass[preprint]{aastex62}
\usepackage{graphicx}% Include figure files
\usepackage{color}%for preprints
\usepackage[T1]{fontenc}
\usepackage{amssymb,amsmath}
\received{xxx, 2018}
\revised{xxx, 2018}
\accepted{xxx, 2018}
\shorttitle{Initial Li Abundances in the Protogalaxy}
\shortauthors{Kusakabe and Kawasaki}
\begin{document}

\title{Initial Li Abundances in the Protogalaxy and Globular Clusters Based upon the Chemical Separation and Hierarchical Structure Formation}

\correspondingauthor{Motohiko Kusakabe}
\email{kusakabe@buaa.edu.cn}

\author[0000-0003-3083-6565]{Motohiko Kusakabe}
\affil{School of Physics, and International Research Center for Big-Bang Cosmology and Element Genesis, Beihang University, \\
  37, Xueyuan Road, Haidian-qu, Beijing 100083, People's Republic of China}

\author{Masahiro Kawasaki}
\affiliation{Institute for Cosmic Ray Research, University of Tokyo, Kashiwa,
Chiba 277-8582, Japan}
\affiliation{Kavli IPMU (WPI), UTIAS, \\
the University of Tokyo, 5-1-5 Kashiwanoha, Kashiwa, 277-8583, Japan}



\begin{abstract}
  The chemical separation of Li$^+$ ions induced by a magnetic field during the hierarchical structure formation can reduce initial Li abundances in cosmic structures. It is shown that cosmological reionization of neutral Li atoms quickly completes as soon as the first star is formed. Since almost all Li is singly ionized during the main course of structure formation, it can efficiently separate from gravitationally collapsing neutral gas. 
  The separation is more efficient in smaller structures which formed earlier. In the framework of the hierarchical structure formation, 
  extremely metal-poor stars can have smaller Li abundances because of their earlier formations. 
  It is found that the chemical separation by a magnetic field thus provides a reason that Li abundances in extremely metal-poor stars are lower than the Spite plateau and have a large dispersion as well as an explanation of the Spite plateau itself.
  In addition,
  the chemical separation scenario can explain 
  Li abundances in NGC 6397 which are higher than the Spite plateau.
  Thus, Li abundances in metal-poor stars possibly keep information on the primordial magnetic field and the structure formation history.
\end{abstract}



\section{Introduction}\label{sec1}
It has been suggested that the primordial magnetic field (PMF) induces a chemical separation of neutral gas and ionized plasma during the structure formation in the early universe \citep[][hereafter KK15]{2015MNRAS.446.1597K}. For specific values of the comoving field amplitude ($\mathcal{O}(0.1)$ nG) and the coherence length corresponding to structures with mass ($10^6 M_\sun$), the chemical separation effectively works. In addition, even when a field amplitude does not have a gradient initially, a gradient is generated during the gravitational contraction of the structure. Therefore, the chemical separation likely occurs if a PMF of sub-nG exists during the cosmological structure formation independent of its inhomogeneity.

At redshifts $z\gtrsim 10$ of the structure formation, the cosmological recombination of Li$^+$ ion has frozen out before completion. Because of a small electron abundance after H recombination \citep{1998A&A...335..403G} and a nonthermal radiation field from a late time recombination of free protons \citep{2005PhRvD..72h3002S}, the relic abundance of Li$^+$ is much larger than that of neutral Li. These abundant Li$^+$ ions as well as protons and electrons can be separated from neutral gas by the magnetic field effect, and escape from structure formation (KK15). In this case, the total lithium abundance relative to hydrogen abundance, i.e., Li/H, in the formed structures can be significantly smaller than the cosmological average value determined at the primordial nucleosynthesis.
When the structure formation proceeds without magnetic field effects, abundances of chemical species heavier than $^1$H are slightly enhanced by a diffusion in thermal structures by less than 1 \% depending on species, structure mass, formation stages, and reionization history \citep{2016MNRAS.459..431M}. The reduction of Li abundance is then much more significant in the case with a magnetic field.

In this letter, we improve this scenario of chemical separation taking into account the cosmological reionization of Li atoms. \citet{2001AJ....122.2850B} have found that cosmological reionization of hydrogen via photoionization \citep[e.g.][]{1967ApJ...147..887W,1985MNRAS.214..137C} occurred in the intergalactic medium (IGM) at the redshift $z \gtrsim 6$ based on the existence of a trough in a spectrum of high $z$ quasi-stellar object (QSO) \citep{1965ApJ...142.1633G}. In Section \ref{sec2}, it is shown that the reionization of Li occurs immediately at the formation of the first star in a volume including the Galactic mass. Since this Li reionization occurs in the beginning of the structure formation, the Li exists in the singly ionized state in the major course of structure formation. As a result, the chemical separation of Li$^+$ ion more effectively reduces the Li abundance in early structures.

Observations of warm dwarf metal-poor halo stars (MPHSs) show a plateau abundance of lithium \citep[e.g.][]{1982A&A...115..357S,2000ApJ...530L..57R,2007A&A...465..587S,2010A&A...522A..26S} called the Spite plateau, which can be interpreted as the primordial abundance \citep{1982A&A...115..357S}. However, this observed abundance at $A(\mathrm{Li})=2.199 \pm 0.086$\footnote{$A(\mathrm{Li})=\log (N_{\rm Li}/N_{\rm H}) +12.$ with $N_i$ the particle number of element $i$.} \citep{2010A&A...522A..26S} is about three times smaller than the abundance derived in standard big bang nucleosynthesis (BBN) model \citep{2016RvMP...88a5004C}. In addition, recent observations of extremely metal-poor (EMP: [Fe/H] <-3\footnote{[A/B]$=\log (N_{\rm A}/N_{\rm B}) -\log (N_{\rm A}/N_{\rm B})_\sun$ where the subscript $\sun$ means the solar value.}) stars show lithium abundances with a large dispersion and an average value below the Spite plateau \citep[e.g.][]{2007A&A...462..851B,2009ApJ...698.1803A,2010A&A...522A..26S,2017PASJ...69...24M}. Furthermore, stellar Li abundances have been measured also in metal-poor globular clusters (MPGCs) (e.g. \citealt{1994A&A...281L..77M,2001A&A...373..905T,2002A&A...390...91B,2006Natur.442..657K,2009A&A...505L..13G} for NGC 6397, \citealt{1995ApJ...452L..13D} for M92, \citealt{1997A&A...322..109P,2007A&A...470..153B} for 47 Tucanae, \citealt{2005A&A...441..549P} for NGC 6752, \citealt{2010A&A...519L...3M} for $\omega$ Centauri, and \citealt{2012A&A...539A.157M} for M4). Although those abundances are similar to the Spite plateau abundance, the Li abundance in NGC 6397, i.e., $A(\mathrm{Li})=2.37 \pm 0.01$ for subgiants \citep{2009A&A...505L..13G}, is slightly larger than the plateau. These observations indicate that some physical process operated and changed the stellar Li abundance during or after the BBN. In Section \ref{sec3}, we give an interpretation of those observations based upon the chemical separation effect caused by the cosmological magnetic field \footnote{It is expected that both isotopes of $^{6,7}$Li$^+$ ions separate similarly although only the $^7$Li$^+$ ion was treated in KK15. The radiative recombination rates and ionization rates of $^{6,7}$Li$^+$ are almost the same due to the nearly equal electric multipole moments in the systems. The frictions on Li$^+$ ions from electrons (equation (63) in KK15) are similar and those from neutral hydrogen \citep[cf.][]{2009PhPl...16e3503K} are expected to change slightly only within a reduced mass factor. Therefore, the chemical separation effectively operates and $^6$Li$^+$ ions move in a strong coupling with electrons. If primordial $^6$Li abundance will be found in observations of MPHSs, it can test the current chemical separation model. However, it is hard to determine the primordial $^6$Li abundance since the abundance in the standard BBN model, i.e., $^6$Li/H$=1.23 \times 10^{-14}$ \citep{2012ApJ...744..158C}, is much lower than the present upper limit at $^6$Li/H$={\mathcal O}(10^{-12})$ \citep{2013A&A...554A..96L} and $^6$Li nuclei are depleted in both pre-main sequence and main sequence phases. }. Our brief conclusion then follows in Section \ref{sec4}.

In this letter, we adopt the natural units for the reduced Planck constant, the Boltzmann constant, and the light speed, i.e., $\hbar=k_{\rm B} =c =1$.

\section{L\lowercase{i} reionization}\label{sec2}
\subsection{Survival of photons with $E_\gamma < E_{\rm H}^{\rm ion}$}\label{sec2a}

Before the cosmological reionization of hydrogen atoms, photons with energies lower than the ionization potential of H, i.e., $E_\gamma <E_{\rm H}^{\rm ion}$, emitted from early astronomical objects such as population III stars and QSOs are not significantly absorbed in the interstellar medium or IGM. Those photons are strongly scattered only in narrow energy ranges at Lyman series transitions of hydrogen, e.g. $E_\alpha=10.20$ eV corresponding to the Ly-$\alpha$ transition. The baryonic matter is mainly composed of hydrogen ($\sim 75$ \% in mass) and helium ($\sim 25$ \%) \citep{2016RvMP...88a5004C}, and they are in the atomic ground states. The ground states of H and He can only absorb continuum photons with energies greater than $E_{\rm H}^{\rm ion} =13.60$ and $E_{\rm He}^{\rm ion} =24.59$ eV, respectively. Therefore, photons with energies $E_\gamma <E_{\rm H}^{\rm ion}$ can escape from the star forming region easily without absorption by H and He, while those with $E_\gamma \geq E_{\rm H}^{\rm ion}$ are destroyed via the photoionization of H (and He). The ionization front of Li around the first star then propagates with the light speed for a long time while those of H and He propagate slower than that \citep[cf.][Appendix]{2012MNRAS.419..873K}.

Abundances of all elements other than H and He are negligible in terms of absorption of ionizing photons since they are smaller than the H abundance by more than 9 orders of magnitude \citep{2012ApJ...744..158C}. As a result, Li ionizing photons with $E_\gamma <E_{\rm H}^{\rm ion}$ can be efficiently used for the Li reionization at the first light from astronomical object which formed via the structure formation. Although those photons are scattered by relic free electrons in the universe, the optical depth is very small as follows.

The destruction rate of ultraviolet photons via Compton scattering before the reionization of H is given by
\begin{eqnarray}
  \Gamma_\gamma^{\rm Com} &=& n_e(z) \sigma_{\rm Th}
  = X n_{\rm b}(z) \chi_{{\rm H}^+} \sigma_{\rm Th} \nonumber \\
  &=& \eta X n_\gamma(z) \chi_{{\rm H}^+} \sigma_{\rm Th} \nonumber\\
  &=& 1.0 \times 10^{-5}~{\rm Gyr}^{-1}
  \left(\frac{\eta}{6.0 \times 10^{-10}}\right)
  \left(\frac{X}{0.75}\right)
  \left(\frac{T_0}{2.7255~{\rm K}}\right)^3
  \left(\frac{1+z}{11}\right)^3
  \left(\frac{\chi_{{\rm H}^+}}{6.5 \times 10^{-5}}\right),\label{eq1}
\end{eqnarray}
where
$n_e$, $n_{\rm b}$, and $n_{{\rm H}^+}$ are the number densities of free electrons, baryons, and protons, respectively,
  $X$ is the primordial mass fraction of hydrogen,
  $n_\gamma =2 \zeta(3) T^3 /\pi^2$ is the number density of background radiation with $\zeta(3) =1.20206$ the zeta function of three and $T=T_0(1+z)$ the photon temperature with $T_0 =2.7255$ K the present photon temperature \citep{2009ApJ...707..916F},
$\eta =n_b/n_\gamma$ is the baryon to photon ratio \citep{2014A&A...571A..16P},
$\chi_{{\rm H}^+}$ is the ratio of the number density of H$^+$ to the total H density \citep{2009A&A...503...47V}, and
$\sigma_{\rm Th} =6.65 \times 10^{-25}$ cm$^{2}$ is the Thomson scattering cross section.
At the first line, we used $n_e=n_{{\rm H}^+}$ from the charge neutrality of the universe noting that the relic He$^+$ abundance is tiny \citep{2009A&A...503...47V}. The scatterings of photons emitted from population III stars with cosmic background electrons are thus very rare, and can be negligible.

This fact that photons with wavelengths longer than the Lyman limit have no important absorber is apparently supported by observed spectra of QSOs \citep{1971ApJ...164L..73L,1972IAUS...44..127L,1998ARA&A..36..267R}. UV continua from QSOs have been observed in which multiple absorption lines corresponding to Ly-$\alpha$ wavelength of hydrogen in absorbers are identified. Except for dense clouds which show absorptions by H and metal lines, the UV photons with $E_\gamma <E_{\rm H}^{\rm ion}$ do not have detectable absorptions during the cosmological Li reionization.

We note that at the Li reionization, the recombination of Li$^+$ in the IGM has already frozen out. Before formation of astrophysical sources for heating IGM starts in the structure formation, the baryonic and electron matter has cooled adiabatically, and the baryonic temperature scales as $T_{\rm g} = 2.3~{\rm K}[(1 + z)/10]^2$ \citep{2004PhRvL..92u1301L}. For example, at the redshift $z=10$ the gas temperature is $T_{\rm g} = 2.8$ K. The recombination rate of Li$^+$ ions\footnote{Here the rate of Li$^+$+$e^-$ reaction only is included. Although the reactions, Li$^+$+H$\rightarrow$Li+H$^+$ and Li$^+$+H$^- \rightarrow$Li+H, also produce the neutral Li, both of the rates are negligibly small. For the former reaction, the detailed balance relation and the forward reaction rate for the final state of the H ground state \citep{1994ApJ...430..435K,1995ApJ...454..545K} lead to the thermal reaction rate of $\langle \sigma v \rangle_{\mathrm{Li}^++\mathrm{H}} =\mathcal{O}(10^{-20})$ cm$^3$ s$^{-1}$ $\exp(-9.52 \times 10^4~{\rm K}/T_{\rm g})$ for $T_{\rm g} ={\mathcal O}(100)$ K. The latter reaction rate \citep{1998A&A...335..403G} is large, i.e., $\langle \sigma v \rangle_{\mathrm{Li}^++\mathrm{H}^-} =6.3 \times 10^{-6} (T_{\rm g}/{\rm K})^{-0.5}$ cm$^3$ s$^{-1}$. However, the recombination rate of Li$^+$ ions is negligibly small because of the tiny abundance of H$^-$, $\chi_{\mathrm{H}^-} \sim 10^{-13}$ at $z =10$.}
is given by
\begin{eqnarray}
  \Gamma_{{\rm Li}^+}^{\rm rec} &=& n_e(z) \langle \sigma v \rangle_{\rm rec} \nonumber \\
%  &=& \eta X n_\gamma(z) \chi_{{\rm H}^+} \langle \sigma v \rangle_{\rm rec} \nonumber\\
  &=& 0.033~{\rm Gyr}^{-1},
%  \left(\frac{\eta}{6.0 \times 10^{-10}}\right)
%  \left(\frac{X}{0.75}\right)
%  \left(\frac{T_0}{2.7255~{\rm K}}\right)^3
%  \left(\frac{1+z}{11}\right)^3
%  \left(\frac{\chi_{{\rm H}^+}}{6.5 \times 10^{-5}}\right)
  %  \left(\frac{\langle \sigma v \rangle_{\rm rec}}{6.4 \times 10^{-11}~{\rm cm}^3~{\rm s}^{-1}}\right),
  \label{eq2}
\end{eqnarray}
where
$\langle \sigma v \rangle_{\rm rec} \approx 1.0 \times 10^{-11}$ cm$^3$ s$^{-1} [T_{\rm g}/(107.7~{\rm K})]^{-1/2}$ is the thermal recombination rate \citep{1998A&A...335..403G}. This recombination rate is significantly smaller than the Hubble expansion rate at $z=10$,
\begin{eqnarray}
  H &\approx& \sqrt{8 \pi G \rho_{\rm m}(z) /3} \nonumber \\
  &=& 1.4~{\rm Gyr}^{-1}
  \left(\frac{\Omega_{\rm m} h^2}{0.14}\right)^{1/2}
  \left(\frac{1+z}{11}\right)^{3/2},
  \label{eq3}
\end{eqnarray}
where $G$ is the Newton constant,
$\rho_{\rm m}$ is the matter density, $\Omega_{\rm m}$ is the matter density parameter, and $h=H_0 /(100~{\rm km~s}^{-1}~{\rm Mpc}^{-1})$ is defined with $H_0$ the Hubble constant.

\subsection{Quick cosmological reionization of L\lowercase{i}}\label{sec2b}

We show that only one massive Pop III star is enough to reionize primordial Li atoms within a volume including Galactic baryon mass. 

Long after the BBN, the first star forms and nucleosynthesis restarts inside the star. It is believed that the initial mass function of early stars was top-heavy, that is, the formation rate of massive stars in the early universe was larger relative to less massive stars.
% than the present rate. 
The upper limit on the first star mass is $\lesssim 100 M_\sun$ \citep{2013RPPh...76k2901B}.

For example, it is assumed that the first massive star has a mass $M_\ast=100 M_\sun$, an Eddington limit luminosity $L_\ast/L_\sun =3.3 \times 10^4 M_\ast/M_\sun$, a blackbody spectrum with the effective temperature $T_\ast=10^5$ K, and the life time $\Delta t_\ast =\eta M_\ast /L_\ast =3.2$ Myr with $\eta=0.00717$ \citep{2017ChPhC..41c0003W} the energy conversion efficiency for nuclear energy via the hydrogen burning \citep{2005ApJ...635..784D}. The solar mass and luminosity are $M_\sun =1.1 \times 10^{57}$ GeV and $L_\sun =3.90 \times 10^{33}$ erg/s, respectively.

The number of ionizing photons for Li emitted from a star is given by
\begin{eqnarray}
  N_\gamma^{\rm Li} &=& \int_{E_{\rm Li}^{\rm ion}}^{E_{\rm H}^{\rm ion}} d E_\gamma \frac{L_\ast(E_\gamma)}{E_\gamma} \Delta t_\ast \nonumber \\
  &=& \frac{120}{\pi^2} \frac{L_\ast \Delta t_\ast}{T_\ast^4}
  \int_{E_{\rm Li}^{\rm ion}}^{E_{\rm H}^{\rm ion}} d E_\gamma \frac{E_\gamma^2}{\exp(E_\gamma/T_\ast) -1} \nonumber \\
%  &=& \frac{120}{\pi^2} \frac{L_\ast \Delta t_\ast}{T_\ast}
%  \int_{E_{\rm Li}^{\rm ion}/T_\ast}^{E_{\rm H}^{\rm ion}/T_\ast} d x \frac{x^2}{{\rm e}^x -1} \nonumber \\
  &=& 6.3 \times 10^{65},
%  \left(\frac{\int_{E_{\rm Li}^{\rm ion}/T_\ast}^{E_{\rm H}^{\rm ion}/T_\ast} d x \frac{x^2}{{\rm e}^x -1}}{0.56} \right)
%  \left( \frac{L_\ast}{3.3 \times 10^6 L_\sun} \right)
%  \left( \frac{\Delta t_\ast}{3.16~{\rm Myr}} \right)
  %  \left( \frac{T_\ast}{10^5~{\rm K}} \right)^{-1},
  \label{eq4}
\end{eqnarray}
where
$L_\ast(E_\gamma) =dL_\ast/dE_\gamma$, and
$E_{\rm Li}^{\rm ion}=5.39172$ eV is the ionization potential of Li.

The number of Li atom in a sphere of the comoving radius $L$ in the universe is given by
\begin{eqnarray}
  N({\rm Li}) &=& \chi_7 X n_{\rm b} (4 \pi L^3 /3)
  %  =\chi_7 X \eta n_\gamma (4 \pi L^3 /3)
  \nonumber \\
  &=& 1.1 \times 10^{58}
  \left(\frac{\chi_7}{5.0 \times 10^{-10}}\right)
  \left(\frac{X}{0.75}\right)
  \left(\frac{\eta}{6.0 \times 10^{-10}}\right)
  \left(\frac{T_0}{2.7255~{\rm K}}\right)^3
  \left(\frac{L}{\rm Mpc}\right)^3,\label{eq5}
\end{eqnarray}
where
$\chi_7$ is the number ratio of Li and hydrogen\footnote{The value in brackets corresponds to the total ratio of Li species which is fixed at the BBN. The actual ratio of neutral Li is, however, much lower than that because of a nonthermal UV field from the late time H recombination \citep{2005PhRvD..72h3002S}.}.

The number of Li atoms in a baryonic material of the mass $M_{\rm b}$ in the universe is given by
\begin{eqnarray}
  N({\rm Li}) &\approx & \chi_7 X M_{\rm b}/m_{\rm H} \nonumber \\
  &=& 4.5 \times 10^{59}
  \left(\frac{\chi_7}{5.0 \times 10^{-10}}\right)
  \left(\frac{X}{0.75}\right)
  \left(\frac{M_{\rm b}}{1.0 \times 10^{12} M_\sun}\right),\label{eq6}
\end{eqnarray}
where
$m_{\rm H} =938.783$ MeV is the mass of hydrogen.
It is seen that only one star is enough to induce the reionization  of Li in a volume much larger than that of the present Galaxy, i.e., $N_\gamma^{\rm Li} \gg N({\rm Li})$.
%If multiple stars form at similar times in the early phase of structure formation in the Galactic volume, the reionization of the region in which the Galaxy forms later proceeds more quickly.
%can be contributed to by more than one star.

\subsection{Cosmological chemical separation of L\lowercase{i}$^+$ ions}\label{sec2c}

The rapid reionization of Li$^+$ occurs at the dawn in the dark age of the universe. Although the ionized degree of Li is expected to be very close to unity even before the reionization \citep{2005PhRvD..72h3002S,2013ARA&A..51..163G}, UV photons produced in the first star would inevitably realize the full ionization independently of any possible deviations in background radiation spectrum and baryonic temperature from those in the standard cosmology. Therefore, during the major part of the structure formation, Li exists in the singly-ionized state. If there is a coherent magnetic field with $B=\mathcal{O}(0.1)$ nG over the scale corresponding to $10^{6} M_\sun$ at the structure formation, Li$^+$ ions as well as protons and electrons can effectively separate from neutral gas which gravitationally collapses (KK15).

Effects of the Li reionization improves understanding of the chemical separation. In KK15, the initial abundances of Li and Li$^+$ were adopted from a chemical reaction network calculation for a homogeneous universe \citep{2009A&A...503...47V}, in which the reionization was not included. The ratio of adopted abundances is Li$^+$/Li$\approx 1$. Then, a significant fraction of Li$^+$ ions could escape from a collapsing structure while neutral Li atoms join the gravitational contraction. The authors suggested that the Li elemental abundance in early structures can be smaller than the cosmological average, i.e., the primordial abundance fixed at BBN, by a factor of two at most. However, when the enhanced reionization rate by the late time H recombination \citep{2005PhRvD..72h3002S} and the reionization by the first star are taken into account, the efficiency of the Li reduction doubles since the ionization degree of Li is not a half but unity.

For example, the chemical separation has been simulated (Case 1 in KK15) for a structure with a mass $M_{\rm str} =10^6 M_\sun$ and a coherent magnetic field $B\sim 0.3$ nG and its gradient over the comoving length $L_0 =10.4$ kpc which completes the gravitational collapse at $z=10$. In that model, about 3/4 of initial Li$^+$ ions escape from the structure formation, and the final average Li abundance in the structure is about 5/8 of the primordial Li abundance. If the Li is singly ionized initially\footnote{The Li reionization by the first star occurs in the early stage of structure formation. The first star forms at $z\sim 20$--30 \citep{2013RPPh...76k2901B}. For example, we take $z=30$, i.e., the cosmic time $t=0.10$ Gyr, which is the early phase of gravitational contraction of structures (cf. Case 1 in KK15). Li ionizing photons with $E_\gamma <E_{\rm H}^{\rm ion}$ quickly propagate to the comoving distance $L_0$ in a time $\Delta t(L_0,z) \approx 0.11$ Myr $(L_0/1~{\rm Mpc}) /[(1+z)/31]$ much shorter than the cosmic expansion time.},
however, the average Li abundance in the structure is 1/4 in this model. This solves the Li problem related to the Spite plateau and the primordial abundance.

\section{Explanation of L\lowercase{i} abundances in MPSs}\label{sec3}

The following three facts have been found from spectroscopic observations for Li in metal-poor stars (MPSs) in the Galaxy on which initial Li abundance is preserved:
(1) such MPSs have the same Li abundance that is known as the Spite plateau with a small dispersion, 
(2) EMP stars have Li abundances the average of which is lower than the Spite plateau and the dispersion of which is much larger than that of the Spite plateau, and
(3) Li abundances in the metal-poor globular cluster (GC) NGC 6397 are higher than the Spite plateau.

These facts can be consistently explained by the chemical separation of Li$^+$ ions during a hierarchical structure formation as follows:

\subsection{L\lowercase{i} abundances in early structures}
Initial Li abundances are lower in smaller structures which formed earlier. First, the chemical separation of Li during the gravitational contraction of a structure is more effective for larger magnetic field intensities and/or its gradient (KK15). Since the PMF is diluted by the cosmic expansion as $B(z) \propto (1+z)^2$, the chemical separation is stronger in the early universe. Second, the chemical separation is only effective for the structure mass of $M_{\rm str} \sim 10^6 M_\sun$ (KK15). This specific mass comes from the condition for a successful contraction of neutral matter and a separation of charged matter from the neutral matter. Thus, as a result of the chemical separation, initial Li abundances are smaller in smaller structures which formed earlier.

In the course of hierarchical structure formation \citep[e.g.][]{1993ApJ...412..455K,1993MNRAS.265..271N}, structures of a given mass can form at different times depending upon the initial degrees of density fluctuation. As a result of continuous structure formation, in the early time, there are many structures with various masses and various formation times. In those structures, Li abundances are different because of different efficiency of the separation of $^7$Li ions from the structure. However, through collisions and mergers of early cosmological structures, gases from different structures mix. The Li abundances in stars which form in structures then gradually approaches to the average value. The Spite plateau abundance is interpreted as this asymptotic Li abundance in a late epoch of structure formation.

In the standard cosmic and Galactic chemical evolution theory, metallicity increases as a function of time reflecting local metal injections from supernovae. Therefore, it is expected that EMP stars tend to form in an early stage of the structure formation before the metallicity increase sufficiently. In such an early time, there exist small structures which formed early. Therefore, the Li abundance can be very small because of strong chemical separation at early times. In addition, the gas mixing via the collisions and mergers of small scale structures has not yet operated. Therefore, if MPSs form in this early epoch, their Li abundances have a large dispersion and they are on average smaller than the Spite plateau abundance.

\subsection{Formations of MPHSs and MPGCs}
In the theory proposed in this work, differences in Li abundances in MPHSs and MPGCs are also explained by times of star formations and histories to formations of their parent bodies. How the Galaxy formed is still not known in detail. Formations of MPHSs and MPGCs then involve large uncertainties.

%A simple scenario for the MPHS formation is production 
A simple scenario is that the MPHSs are formed in small structures which are building blocks of the Galaxies. After the stars formed, those building blocks are engulfed in the proto-Galaxy. Then, the stars survive in the Galaxy and are observed today as MPHSs, while gases later form a Galactic disk and disk stars would form there. In such a scenario, MPHSs form before the mergers of their parent bodies with the proto-Galaxy.

On the other hand, although the formation of MPGCs is uncertain still, there are possible scenarios. They include: (1) production in small structures at star bursts before their mergers to the proto-Galaxy \citep[e.g.][]{2012MNRAS.421L..44B,2014MNRAS.444.3670P}, and (2) production at the mergers from mixture of gases (\citealt{1978ApJ...225..357S,1992ApJ...384...50A} and see \citealt{2009PASJ...61..481S} for simulations of star formations at galaxy mergers).

%\begin{itemize}
%\item Case (1):
  In case (1), Li abundances of MPGCs are determined by histories of the host structures of MPGCs. If those stars formed at relatively late times of the structure formation, the parent body has a large Li abundance because of less effective chemical separation.

%\item Case (2):
  In case (2), at mergers of small structures into the proto-Galaxy, GCs possibly form after a mixing of gases from the merging structures and the proto-Galaxy. Galactic gas is relatively Li-rich excepting small structures inside the Galaxy since the chemical separation of Li is effective only in small structures (KK15). Therefore, the mixing of gases of the small structures and proto-Galaxy may result in large Li abundance in mixed gas even if the Li abundances in the merging small structures are small.
%\end{itemize}

In these two cases, the Li abundances in MPGCs can be both higher than the Spite plateau as observed in NGC 6397 and similar to the plateau, depending on the origin of the MPGCs.

\begin{figure}[ht!]
  \epsscale{1.}
  \plotone{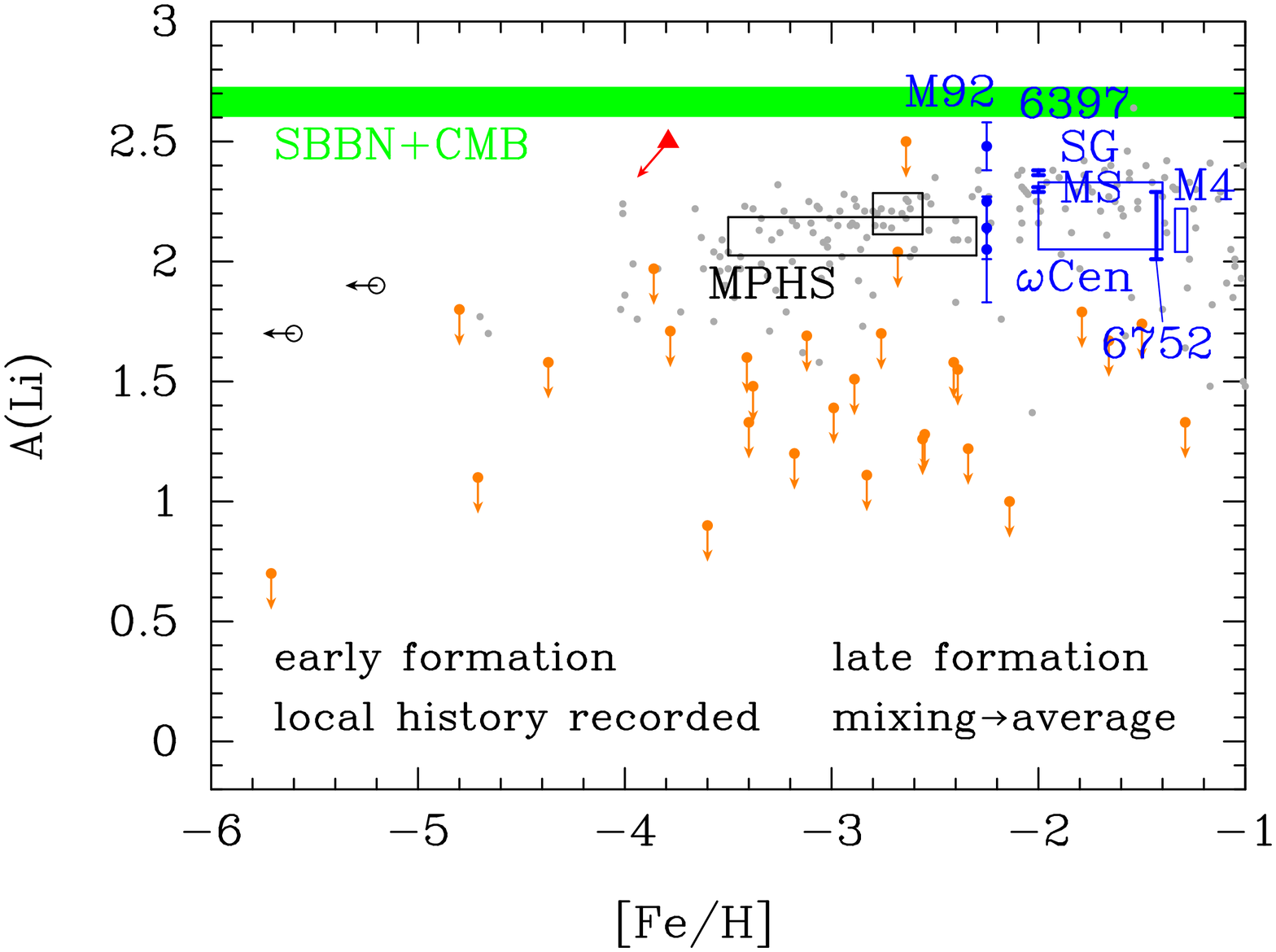}
  \caption{Li abundances as a function of metallicity taken from \citet[][and references therein]{2017AJ....154...52M}, \citet{2018A&A...612A..65B,2019ApJ...871..146F}. Symbols correspond to stars with detections of Li and Fe abundances (dots), only upper limits on Li abundances (solid circles), only upper limits on Fe abundances (open circles), and upper limits on both Li and Fe abundances (a triangle), respectively. Boxes show the Spite plateau of MPHSs (\citet{2000ApJ...530L..57R} (wider one) and \citet{2010A&A...522A..26S} (narrower one)), and abundances of $\omega$ Centauri \citep{2010A&A...519L...3M} and M4 \citep{2012A&A...539A.157M}, while dots with error bars show abundances of M92 \citep{1995ApJ...452L..13D}, NGC 6397 (the main sequence (MS) and subgiants (SG)) \citep{2009A&A...505L..13G} and NGC 6752 \citep{2005A&A...441..549P}, as labeled. The horizontal band is located at the standard BBN prediction of primordial Li abundance \citep{2016RvMP...88a5004C}.}
\label{fig:li_abundance}
\end{figure}

Figure \ref{fig:li_abundance} shows stellar Li abundances as a function of [Fe/H].
In the current scenario, MPHSs with very low [Fe/H] values form early and have low Li abundances because of the effective chemical separation. In addition, a large dispersion in the Li abundances is expected since the mixing of gases originating from different structures is incomplete and an inhomogeneity in the abundances remains. Stellar Li abundances approach to an average value as the structure formation proceeds and the [Fe/H] increases.

\section{Conclusions}\label{sec4}

The efficiency of the chemical separation of Li$^+$ ions by a postulated PMF with comoving intensities of $\sim$ nG is enhanced in the early epoch of the structure formation. As soon as the first star begins to emit light, neutral Li atoms in the volume including the Galaxy mass are easily reionized. As a result, almost all Li nuclei can join the chemical separation, and Li abundances inside collapsing structures can be significantly smaller than the primordial abundance. This explains that Li abundances in MPSs are smaller than the primordial abundance.

Lithium abundances in EMP stars have an average value lower than the Spite plateau and a dispersion larger than that of the plateau. This trend is possibly explained since the chemical separation works more effectively in the early epoch when the amplitude of PMF is larger. Through collisions and mergers during the hierarchical structure formation, gases with different Li abundances mix and an asymptotic abundance is expected to be realized. That asymptotic abundance corresponds to the Spite plateau. Thus, MPSs with [Fe/H]$\gtrsim -3$ have the similar Li abundance since they formed in a later epoch of structure formation.

The Li abundances in NGC 6397 are higher than the Spite plateau. In the current model, this indicates that the parent body of NGC 6397 are not affected by the chemical separation very much. The high Li abundance is possible if the formation of the parent body effectively progressed late or the PMF penetrating the body was initially weaker than the average value in the Galactic volume.

\acknowledgments

We are grateful to Michiko S. Fujii for helpful information on scenarios of globular cluster formations and to Tadafumi Matsuno for providing compiled data of Li abundances.
The work of M. Kusakabe was supported by NSFC Research Fund for International Young Scientists (11850410441). The work of M. Kawasaki was supported by JSPS KAKENHI Grant Nos. 17H01131 and 17K05434, MEXT KAKENHI Grant No. 15H05889, World Premier International Research Center Initiative (WPI Initiative), MEXT, Japan.

\end{document}